\documentclass[12pt,preprint]{aastex}

\shorttitle{SN associated with GRB 091127}
\shortauthors{Cobb et al.}

\begin{document}

\title{Discovery of SN 2009nz Associated with GRB 091127}

\author{B.~E. Cobb\altaffilmark{1}, J.~S. Bloom\altaffilmark{1}, D.~A. Perley\altaffilmark{1}, A. N. Morgan\altaffilmark{1}, S.~B. Cenko\altaffilmark{1}, and
A.~V. Filippenko\altaffilmark{1}}
\email{bcobb@astro.berkeley.edu}

\altaffiltext{1}{Department of Astronomy, University of California,
  Berkeley, CA 94720-3411}

\begin{abstract}

We report SMARTS, Gemini and \textit{Swift}-UVOT observations of the optical transient (OT)
associated with gamma-ray burst (GRB) 091127, at redshift 0.49, taken
between 0.9 hr and 102 days following the \textit{Swift} trigger.  In
our early-time observations, the OT fades in a manner
consistent with previously observed GRB afterglows.  However, after 9
days post-burst, the OT is observed to brighten for a period of
$\sim2$ weeks, after which the source resumes fading.  A comparison
of this late-time ``bump'' to SN 1998bw (the broad-lined Type Ic
supernova associated with GRB 980425), and several other GRB
supernovae (SNe), indicates that the most straightforward explanation is
that GRB 091127 was accompanied by a contemporaneous SN (SN 2009nz) that
peaked at a magnitude of $M_V=-19.0\pm0.2$.  SN 2009nz is
globally similar to other GRB supernovae, but evolves slightly faster
than SN 1998bw and reaches a slightly dimmer peak magnitude.  We also
analyze the early-time UV-optical-IR spectral energy distribution of
the afterglow of GRB 091127 and find that there is little to no
reddening in the host galaxy along the line-of-slight to this burst.

\end{abstract}


\section{Introduction}

The list of gamma-ray bursts (GRBs) that are associated with
supernovae (SNe) has grown in the last decade (e.g., GRB 030329/SN
2003dh --- \citealt{Hjorth+03}, \citealt{Stanek+03},
\citealt{Bloom+04}; GRB 031203/SN 2003lw --- \citealt{Cobb+04},
\citealt{Gal-Yam+04}, \citealt{Malesani+04}, \citealt{Thomsen+04}; GRB
060218/SN 2006aj --- \citealt{Campana+06}, \citealt{Ferrero+06}, \citealt{Mirabal+06}, 
\citealt[][and references therein]{Cobb+06a}; GRB 050525a ---
\citealt{DellaValle+06a}; see also \citealt{Woosley+Bloom06}),
indicating that at least some (perhaps most) long-duration GRBs result
from the core collapse of massive stars.  However, as evidenced by the
failure to detect SNe associated with two nearby long-duration GRBs
(GRB 060505 at redshift $z=0.0889$ and GRB 060614 at $z=0.125$:
\citealt{Cobb+06b}, \citealt{DellaValle+06b}, \citealt{Fynbo+06},
\citealt{Gal-Yam+06}, \citealt{Xu+09}), we do not yet have a complete
understanding of the progenitors of long-durations GRBs and the
apparent origin of the diversity of their associated SN properties.
Long-duration GRB 091127 ($T_{90} = 7.1$\,s, \citealt{gcn10191}),
presented us with a new opportunity to investigate the GRB-SN
connection.

GRB 091127 triggered the \textit{Swift} Burst Alert Telescope
\citep{Gehrels+04} on 2009 Nov. 27 at 23:25:45 (UT dates are used
throughout this paper; \citealt{gcn10191}).  The prompt emission
had a power-law index of $2.05\pm0.07$ ($-0.4$ to $7.5$ s post-burst)
and a fluence of $9.0\pm0.3\times10^{-6}$ erg cm$^{-2}$ (15-150 keV) \citep{gcn10197}.   
The \textit{Swift} X-ray Telescope observed the X-ray afterglow with
a photon spectral index of $1.98^{+0.15}_{-0.14}$ and absorption column
of $9.8^{+3.3}_{-3.1}\times10^{20}$ cm$^{-2}$ \citep{gcn10201}. 

\textit{Swift} could not
immediately slew to the burst due to Earth-limb constraints, but an
optical transient (OT) was quickly identified using the robotic 2-m
Liverpool Telescope \citep{gcn10192}.  \textit{Swift} Ultraviolet and
Optical Telescope (UVOT, \citealt{Roming+05}) observations, beginning
$\sim 50$ min post-burst, confirmed the existence of this OT at
RA$_{\rm J2000}$ = $02^\mathrm{h}26^\mathrm{m}19.89^\mathrm{s}$, Dec$_{\rm J2000}$ =
$-18^\circ57^\prime08\farcs5$ \citep{gcn10193}.

Gemini-North Multi-Object Spectrograph (GMOS, \citealt{Hook+04}) and
Very Large Telescope X-shooter spectra of the OT revealed emission
features from the underlying host galaxy of the GRB at $z=0.49$
\citep{gcn10202,gcn10233}.  Because of this relatively low redshift,
GRB 091127 was an excellent candidate for follow-up observations in
search of a GRB-related SN.

In this {\it Letter}, we present optical data that show a late-time
rebrightening in the OT of GRB 091127.  We interpret this extra
component of light as being due to a SN (SN 2009nz, \citealt{cbet2288}) associated with GRB 091127.
The observations, data reduction, and photometry are reported in \S 2.
In \S 3, we consider each component of light associated with the OT
and note the absence of reddening in the burst's host galaxy.  We
conclude in \S 4 with a comparison of past GRB-SNe.  Throughout this
paper, we assume the standard cosmological model with $\Omega_\Lambda=
0.73$, $\Omega_m= 0.27$, and a Hubble constant of 71 km s$^{-1}$
Mpc$^{-1}$.

\section{Observations, Data Reduction and Photometry}
\subsection{SMARTS Optical/IR Observations}

We began observing the field of GRB 091127 on 2009 Nov. 28 at 01:15,
$\sim1.8$ hr post-burst \citep{gcn10191}, using ANDICAM (A Novel Dual
Imaging CAMera) mounted on the Small and Moderate Aperture Research
Telescope System (SMARTS)\footnote{http://www.astro.yale.edu/smarts.}
1.3-m telescope at Cerro Tololo Inter-American
Observatory.\footnote{http://www.astronomy.ohio-state.edu/ANDICAM.}
Initial SMARTS observations were obtained in short exposures over
ANDICAM's full wavelength range (\textit{BVRIJHK}); as the OT faded,
longer observations were obtained with fewer filters.  A thorough
analysis of the afterglow of GRB 091127, including the full-color
SMARTS dataset, is available in Troja et al. (2010, in prep.).  Over
11 epochs between 1.8 hr and 24 days post-burst, dithered images were
obtained and then reduced and combined using standard IRAF\footnote{IRAF is distributed by 
National Optical Astronomy Observatory, which is operated by the Association for Research in 
Astronomy, Inc., under cooperative agreement with the NSF.} 
tasks (see Table 1, Fig. 1).

The brightness of the OT of GRB 091127 was measured using
seeing-matched aperture photometry relative to a set of on-chip,
nonvariable sources.  Relative magnitudes were converted to apparent
magnitudes by comparison, on a photometric night, with Rubin 149
Landolt standard stars \citep{L92}.  In addition to the relative
measurement error, there is a systematic error of 0.05 mag associated
with the uncertainties in this photometric calibration.  All
photometry in this paper is corrected for a Galactic reddening of
$E_{B-V}=0.038$ mag \citep{Schlegel+98}.

\subsection{Gemini Optical Observations}

We obtained images of GRB 091127 using GMOS on the 8-m Gemini-South
telescope.  Five epochs of GMOS $i'$-band imaging were obtained
between 9 and 102 days post-burst (see Table 1).  Each set of Gemini
images consisted of dithered exposures reduced and combined using the
standard \texttt{gemini.gmos} IRAF package.

Seeing-matched, relative aperture photometry was performed on the OT
of GRB 091127. The relative to apparent magnitude transformation
utilized two stars common to both the SMARTS and Gemini images.  The
$i'$-band apparent magnitudes of the stars were determined with the
SDSS transformation equations of
\cite{Jordi+06}\footnote{http://www.sdss.org/dr7/algorithms/sdssUBVRITransform.html\#Jordi2006.},
utilizing the stars' $I$- and $R$-band SMARTS magnitudes.  These
Gemini $i'$-band magnitudes were then transformed back into the
$I$-band to match the SMARTS photometric system.  While these
transformations may introduce some systematic error into the Gemini
photometry, they do not affect the relative magnitudes.  The match
between SMARTS and Gemini values measured at similar epochs suggests
that no significant error has been introduced.
 
ISIS \citep{Alard00} kernel-convolved image subtraction was carried
out on the Gemini images (Fig. 2).  The image obtained 102 days
post-burst was used as the subtraction reference frame.  Residual
light is evident in each subtracted frame, indicating that the OT was
dimmest in the final image.  This is expected, of course, if the
earlier images contain afterglow light.  However, both the image
subtraction and the photometry indicate that the transient
\textit{brightens} by $0.06\pm0.02$ mag between 9 and 18 days
post-burst.  As described in \S3, we interpret this brightening to
indicate the presence of an underlying GRB-SN.

\subsection{UVOT Observations}

UVOT observations of GRB 091127 began on 2009 Nov. 28 at 00:19:29,
53.65 min after the trigger, and the afterglow was detected in all
utilized filters.  Count rates were measured using
$5''$ apertures on data taken with the $U$, $UVW1$, and $UVM2$
filters, and were calibrated using the UVOT Photometric System
described by \cite[][see Table 1]{Poole+08}.

\section{Data Analysis}

The OT of GRB 091127 consists of three distinct sources of light.  The
first source is the steady contribution from the underlying host
galaxy, which dominates at late times ($t_\mathrm{obs}\gtrsim50$ days). The
second source, most important at early times ($t_\mathrm{obs}\lesssim6$
days), is the decaying optical afterglow (OAG) of the GRB.  The third
source is the rising and then decaying light from the putative SN
associated with GRB 091127, SN 2009nz, which begins to contribute significantly
to the system several days post-burst.  Below, we consider each of
these contributions in turn.

\subsection{Host Galaxy}

We assume that in the final Gemini image ($\sim102$ days post-burst),
all transient sources associated with GRB 091127 no longer contribute
significantly to the observed optical flux.  Therefore, we take the
magnitude at that epoch ($I=22.54\pm0.05$ mag) to be the brightness
of the host.  If a small contribution of extra light from the OAG or
SN is still present in our images at this epoch, then we may have
overestimated the brightness of the host; thus, we will have
oversubtracted the host contribution and slightly underestimated the
peak brightness of the GRB-SN associated with GRB 091127.  We estimate
that the systematic uncertainty in the SN peak brightness introduced
by host-galaxy subtraction is $<0.1$ mag.

\subsubsection{GRB Position}

The host galaxy of GRB 091127 exhibits a nonstellar radial profile
(full width at half-maximum intensity $\sim$1\farcs0 when the seeing
was 0\farcs7), with a slight elongation from northwest to southeast.
The centroid of the OT is offset from the host-galaxy center by
0\farcs$09\pm0$\farcs01 west and 0\farcs$26\pm0$\farcs01 south.
The position of the GRB is, therefore, inconsistent with the center of
the galaxy.  At the distance of the host ($z=0.49$),
$1^{\prime\prime}$ is 6.02 kpc in projection.  Thus, the GRB occurred
$\gtrsim1.66\pm0.06$ kpc from the host center, which is a typical
offset for a long-duration GRB \citep{Bloom+02a,Fruchter+06}.

\subsubsection{Host-Galaxy Reddening}

SMARTS observations were obtained with a cadence designed to ensure
that the final combined frames in each filter are referenced to the
same mid-exposure time.  The \textit{BVRIJHK} images obtained during
the first SMARTS epoch have a common mid-exposure time of 2.2 hr
post-burst.  The OAG was brightest during this epoch, so we use these
data to build the spectral energy distribution (SED) of the afterglow
in order to evaluate whether there is significant extinction along the
line-of-sight through the GRB host galaxy.

To help constrain the host reddening, we extend the SED blueward of
the SMARTS \textit{B}-band filter using \textit{Swift}-UVOT
ultraviolet observations.  The UVOT light curve was best sampled in
the $U$ filter, and a power-law fit to the data yield a decay slope of
$\alpha=0.55\pm0.05$ (for $f_\nu\propto t^{-\alpha}$, where
$f_\nu$ is the transient's flux density and $t$ is the time since the
burst trigger), consistent with the decay rates inferred from SMARTS
\citep{gcn10244} and SkyNet/PROMPT \citep{gcn10219} observations at
similar times.  Assuming this decay rate, UVOT magnitudes were
extrapolated to the common time of 2.2 hr post-burst ($U=16.45\pm0.10$ mag, $UVW1=16.22\pm0.10$ mag, 
$UVM2=16.08\pm0.10$ mag) in order
to match the SMARTS epoch.

Figure 3 shows the SMARTS/UVOT SED of the OAG of GRB 091127.  After
correction for Galactic extinction, the observed UV-optical-IR SED was
fit assuming an intrinsic power law affected by an extinction screen
at the host redshift of 0.49.  We examined several models, including
Milky Way-like extinction, SMC-like extinction, and LMC-like
extinction using the parameterization of \citet{Fitzpatrick99}. The
data were also fit to an unextinguished power law.  Fits with a small
amount of host-galaxy extinction were statistically acceptable, as was
the fit to the unextinguished case.  Given that the addition of
host-galaxy extinction does not significantly improve the model fit,
we suggest there is little to no significant extinction along the
line-of-sight to GRB 091127.  The $3\sigma$ upper limit on host-galaxy
extinction is $A_V<0.5$ mag.  If a small amount of extinction is
present in the host galaxy, we will slightly underestimate the peak
brightness of the SN associated with GRB 091127.

\subsection{Optical Afterglow}

The decay of the OAG of GRB 091127 is modeled by a broken power law.
During the first three epochs, the power-law decay index calculated
from the SMARTS $I$-band observations is $\alpha = 0.54\pm0.02$.  At
$\sim 0.3$ days post-burst, the power law steepens to $\alpha =
1.29\pm0.03$.  This decay index is similar to that reported in other
optical filters \citep{gcn10219}.  Observations taken up to 6 days
post-burst are dominated by the OAG.  The transient's behavior in
later epochs, however, deviates significantly from this power-law
decay.

This late-time deviation may, in part, be due to light from the
underlying host galaxy, so we subtract the host-galaxy contribution
(see Table 1).  At early times, this subtraction has very little
effect on the brightness of the transient, but it does slightly
steepen the later-time decay rate of the afterglow to $\alpha =
1.52\pm0.03$.  Despite this subtraction, the transient still does not
behave as expected for a very late-time OAG.  Instead of following a
power-law decay, the transient \textit{brightens}.

It is not uncommon for an afterglow to brighten, though this usually
occurs during its early-time evolution during or immediately after
the prompt phase of the GRB \citep{Oates+09,Kann+10}.  Optical flaring is also relatively common and
is generally attributed to refreshed shocks or reverse shocks
\citep[e.g.,][]{Greiner+09}. The OAG of GRB 091127 shows no
indications of significant brightening or flaring at early times
\citep{gcn10219}.  The X-ray afterglow of GRB 091127 also shows no
flaring activity at early times and does not deviate from a simple
power-law decay when the OT brightens.  The brightening occurs several
weeks post-burst and, therefore, cannot be easily attributed to GRB
central-engine activity.  When a late-time rebrightening with a
similar timescale to that of GRB 091127 has been observed following a
GRB (see \citealt{Woosley+Bloom06} and references therein), the light
curve ``bump'' has been attributed to a SN that occurred concurrently
with the GRB.

\subsection{SN 2009nz Associated with GRB 091127}

To examine the late-time brightening component of GRB 091127, both the
host galaxy and OAG contributions are subtracted (see Table 1, Fig.
1, \textit{inset}).  To account for possible errors, a range of
subtractions was implemented assuming every possible permutation of a
$\pm3\sigma$ error on the host brightness and the afterglow model.
This $3\sigma$ confidence region is shown as the gray area in the
inset of Figure 1 and likely overestimates the actual errors on the SN
associated with GRB 091127.  Variations in the assumed host-galaxy
magnitude most strongly affect the brightness of the late-time
observations, while changes in the fit to the afterglow alter the
earlier-time observations.

Regardless of the exact values assumed for the host galaxy and OAG
subtraction, the classic rise and then decay of a SN is clear.  The
observed peak magnitude of SN 2009nz is
$I=22.3\pm0.2$ mag, which occurs at $22\pm3$ days post-burst.  At the
burst's redshift of 0.49, this is equivalent to an absolute peak
magnitude of $M_V=-19.0\pm0.2$ occurring at a rest-frame time of
$15\pm2$ days post-burst.  
For these calculations, we have assumed that the K-correction of
this GRB-SN is similar to that of the prototypical GRB-SN, SN 1998bw,
and employ a time-dependent, generalized K-correction \citep{Kim+96}
that utilizes the spectra of SN 1998bw from \cite{Patat+01} and the
photometry of \cite{Galama+98}.

The observed $I$-band light curve of SN 1998bw at $z=0.49$ is shown as
a curve in the inset of Figure 1.  The SN associated with GRB 091127
evolves faster than SN 1998bw, and reaches a slightly dimmer peak
magnitude.  This peak magnitude could be brighter if the
SN has been reddened by its host galaxy; the $3\sigma$ upper limit on
reddening is $A_V<0.5$ mag and, therefore, extinction is unlikely to
alter the peak magnitude by more than a few tenths of a magnitude.

Given the resemblance between SN 1998bw and the extra component of
late-time light in GRB 091127, alternatives to the SN explanation for
the source of this light are difficult to support.  While it was
initially speculated that some late-time GRB optical afterglow
rebrightenings might be attributed to ``dust echos''
\citep[e.g.,][]{WD00}, later analysis concluded that these models
could not fit the data and SNe were much more natural explanations
\citep{Reichart01}.  Furthermore, similar late-time ``bumps'' in the
OAG light curves of other GRBs (e.g., GRB 021211,
\citealt{DellaValle+03}; GRB 050525a, \citealt{DellaValle+06a}) have
been shown spectroscopically to be GRB-SNe.  Hence, we consider our
observations to be an extremely strong photometric case for a SN
associated with GRB 091127.
  
\section{GRB-SNe Comparison}

We compare the absolute $V$-band light curve of SN 2009nz with other GRB-SNe whose SN light curves can be separated
from their OAGs (SN 1998bw/GRB 980425, SN 2003dh/GRB 030329, SN
2003lw/GRB 031203, and SN 2006aj/GRB 060218; see Fig. 4).  The GRB-SNe
are globally very similar in terms of rise times and peak magnitudes.
The GRB-SNe cluster fairly tightly in peak brightness, though SN
2003lw appears to be somewhat brighter than the others.  The exact
peak magnitude of SN 2003lw, however, depends on a large and uncertain
amount of Galactic and host-galaxy extinction
\citep[e.g.,][]{Malesani+04}.  Depending on the reddening values
assumed, SN 2003lw may be up to 0.5 mag dimmer than shown in Figure 4,
thus making its peak brightness more in line with the other GRB-SNe.
The light curve of SN 2003dh is also subject to some uncertainty
because of the difficulty of separating the SN component of GRB 030329
from its very bright OAG \citep{Deng+05}.

A significant variation among the GRB-SNe is their rise times, with SN
2006aj peaking the fastest and SN 2003lw taking the longest time to
peak.  There appears to be a trend toward brighter GRB-SNe evolving
more slowly than fainter GRB-SNe \citep{Bloom+02b}.  For every 0.1 mag
of dimming (brightening) compared to the peak brightness of SN 1998bw,
the GRB-SNe evolve $6\pm2\%$ more quickly (slowly). This trend,
however, has an unknown amount of associated error given
the uncertainties associated with the light curves of SN 2003lw and SN
2003dh and the relative sparsity of $V$-band data points for SN
2003lw and, therefore, may not be significant (see \citealt{Ferrero+06}).

Considering the similarities in the light curves of SN 1998bw and SN 2009nz, it is likely that they ejected
comparable amounts of ${^{56}}$Ni (0.5--0.7 M$_\odot$, see
\citealt{Woosley+Bloom06} and references therein).  The similarities between these two SNe
are particularly interesting because of the large disparity between
the gamma-ray energy associated with GRB 980425 and GRB 091127.  GRB
980425 was a particularly subluminous GRB, with $E_{\rm iso} \approx
10^{48}$ ergs, while GRB 091127 was a much more energetic burst with
$E_{\rm iso} \approx 10^{53}$ ergs \citep{gcn10197}.  Even if GRB 091127 was highly
collimated, the gamma-ray energy output of GRB 091127 corrected for
beaming is still at least a few orders of magnitude larger than that
of GRB 980425.  While there is evidence of a large population of
local, low-energy, long-duration GRBs without significant OAGs that
are associated with SN 1998bw-like SNe (see \citealt{Cobb+06a}), GRB
091127 with its large $E_{\rm iso}$ and bright OAG is not a member of
this class.  Instead, GRB 091127 is much more similar to GRB 030329
(associated with SN 2003dh).  The addition of another member to the
class of GRB-SNe associated with ``typical'' energy, cosmological GRBs
provides strong supporting evidence that many, if not most, long-duration GRBs are produced by
the core collapse of massive stars.

\acknowledgments

We thank J.~Espinoza, A.~Miranda, and S.~Tourtellotte for
assistance with SMARTS observations (NOAO programs 2009B-0469/2010A-0113) and data reduction.
B.E.C. acknowledges NSF Astronomy \& Astrophysics
Postdoctoral Fellowship AST-0802333.  A.V.F. and S.B.C. acknowledge
generous support from Gary \& Cynthia Bengier, the Richard \& Rhoda
Goldman Fund, NASA/\textit{Swift} grants NNX09AL08G/NNX10AI21G,
and NSF AST-0908886.  A.N.M. acknowledges support from an NSF
Graduate Research Fellowship. The Gemini Observatory (data
obtained under programs GS-2009B-Q-5/GS-2010A-Q-5) is operated by
AURA under an agreement with the NSF on behalf of the Gemini
partnership: the NSF (US), the Science and Technology Facilities
Council (UK), the National Research Council (Canada), CONICYT (Chile),
the Australian Research Council (Australia), Minist\'{e}rio da
Ci\^{e}ncia e Tecnologia (Brazil), and Ministerio de Ciencia,
Tecnolog\'{i}a e Innovaci\'{o}n Productiva (Argentina).  We
acknowledge the use of public data from the \textit{Swift} archive.

 \clearpage

\begin{figure}
\includegraphics[width=1\textwidth]{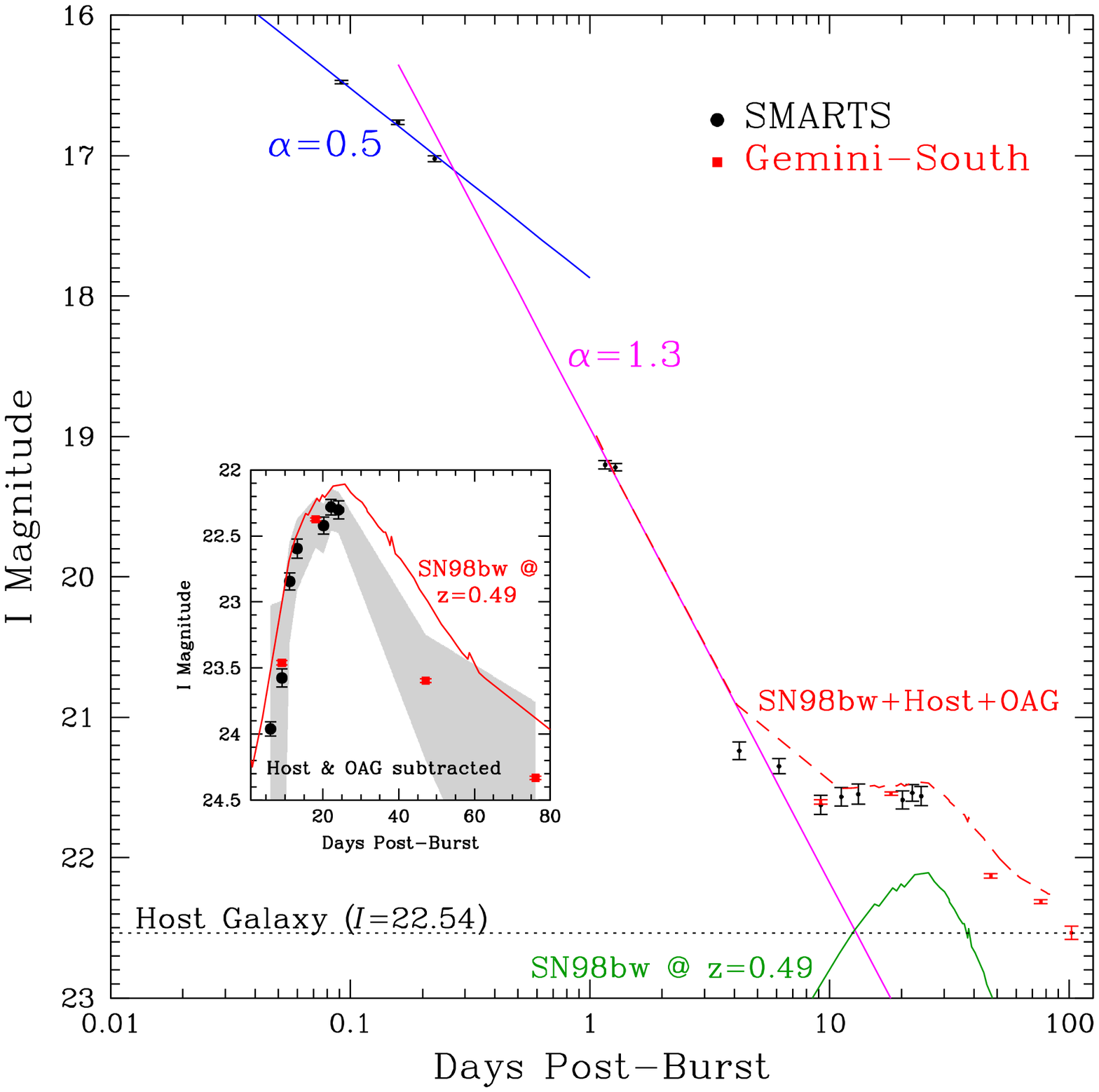}
\caption{Observed $I$-band light curve of GRB 091127's OT.  
  The initial decay rate is $\alpha\approx0.5$, which
  steepens to $\alpha\approx1.3$ ($\alpha\approx1.5$ after
  host-galaxy subtraction).  The brightness of SN 1998bw at $z=0.49$
  is shown for comparison.  Also shown is the light curve expected if
  a SN 1998bw-like SN had been associated with this burst's host and
  OAG (SN98bw+Host+OAG) --- this is brighter than the observed points,
  indicating that the GRB 091127 SN is slightly dimmer than SN 1998bw.
  \textit{Inset:} An observer-frame comparison of SN 1998bw and the SN
  associated with GRB 091127.  The points show the SMARTS/Gemini
  observations with both the host and the OAG contributions removed.
  The grey shaded region is the $3\sigma$ confidence region, allowing
  for possible errors in the magnitude of the host and the afterglow
  model.  This SN evolves somewhat more quickly than SN 1998bw and
  peaks at a slightly dimmer magnitude.}
\end{figure}

\begin{figure}
\includegraphics[width=1\textwidth]{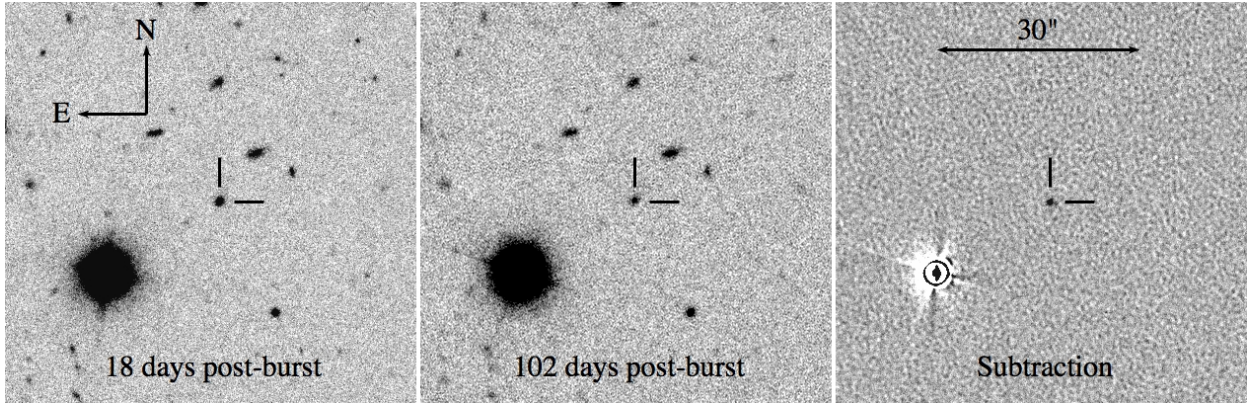}
\caption{Gemini $i'$-band images of GRB 091127 taken at
  18 days (\textit{left}) and 102 days (\textit{center}) post-burst.
  The position of the transient/host galaxy is
  indicated.  The \textit{right panel} shows the image
  subtraction of the two previous panels.  Residual light is
  evident in the subtracted frame and is comprised of
  light from both the OAG and SN 2009nz.}
\end{figure}

\begin{figure}
\includegraphics[width=1\textwidth]{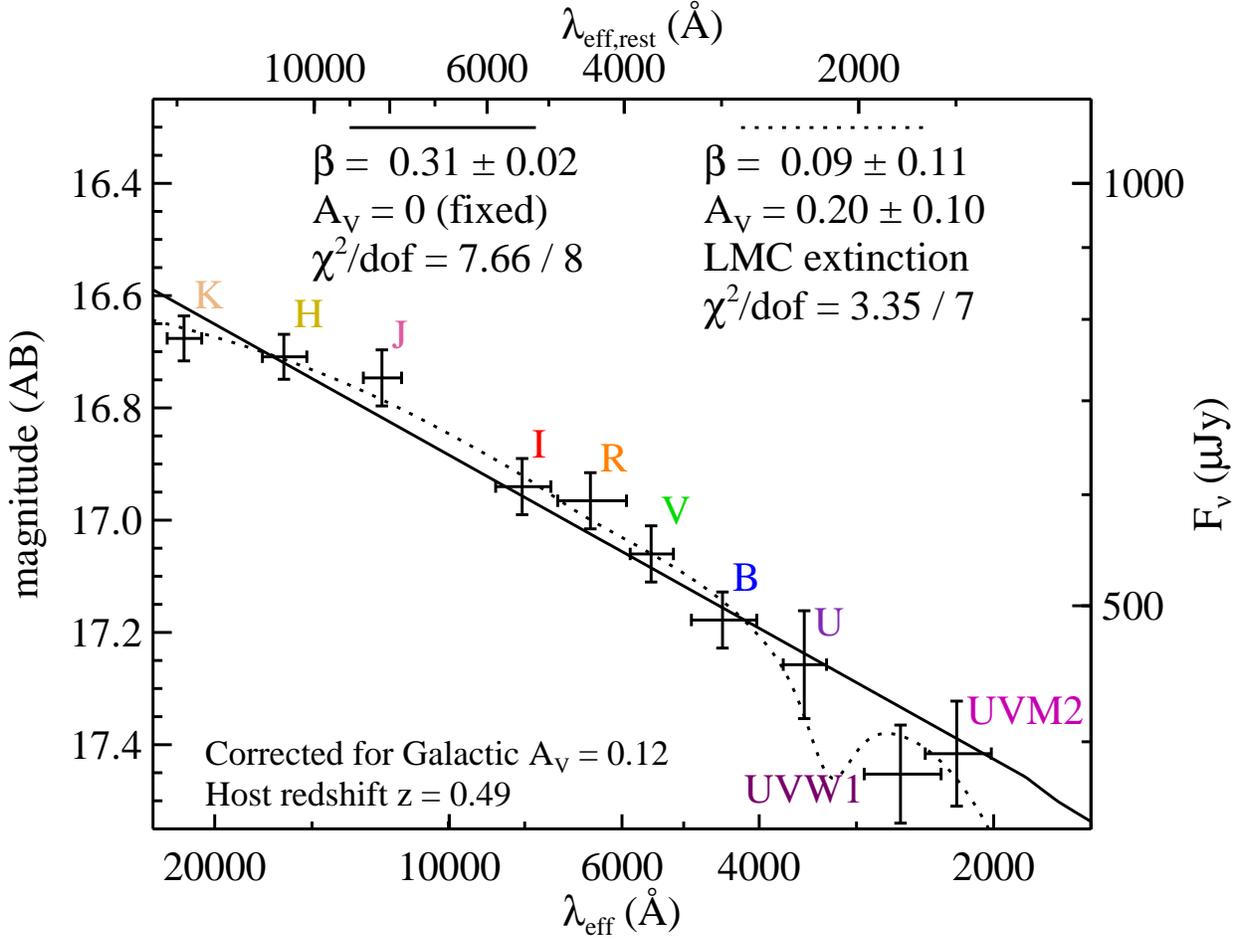}
\caption{UV-optical-IR SED of the OAG of GRB 091127 at $\sim2.2$ hr
  post-burst (corrected for Galactic extinction).  The solid line
  shows a power-law fit assuming no host-galaxy extinction.  The
  dotted line shows a fit using a model that includes a small amount
  of LMC-like extinction in the host.  Both fits are statistically
  acceptable, indicating that there is little to no significant 
  host-galaxy extinction along the line-of-sight to this GRB.}
\end{figure}

\begin{figure}
\includegraphics[width=1\textwidth]{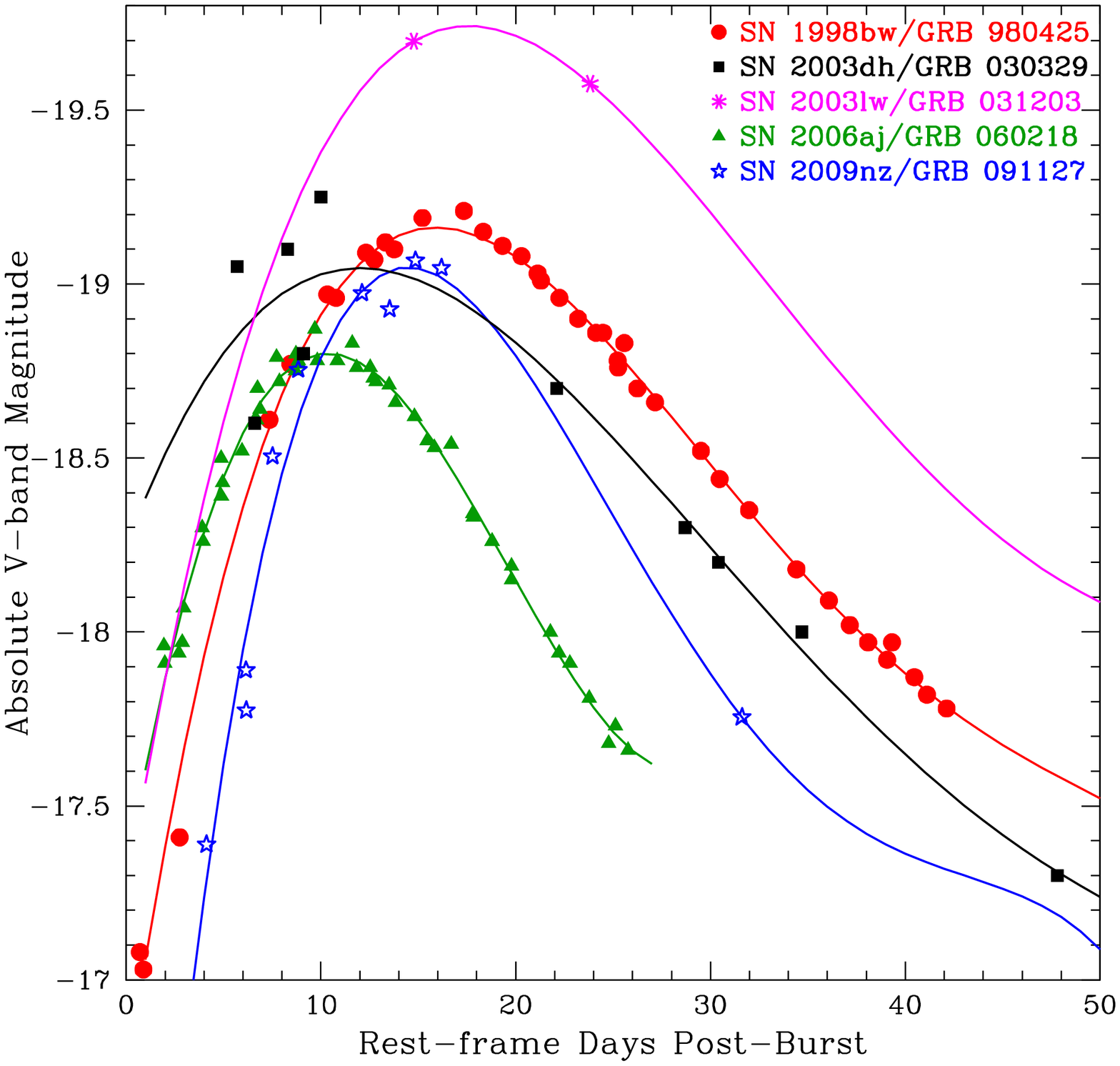}
\caption{Comparison of GRB-SNe absolute $V$-band light curves, with
  data obtained from the following references: SN 1998bw,
  \citealt{Galama+98}; SN 2003dh, \citealt{Deng+05}; SN 2003lw,
  \citealt{Malesani+04}; SN 2006aj, \citealt{Mirabal+06}, and
  \citealt{Ferrero+06}.  To guide the eye, the points have been fit
  with simple polynomial curves.  GRB 031203 occurred behind a large
  and uncertain amount of Galactic and host-galaxy extinction and
  could, therefore, be as much as 0.5 mag dimmer.}
\end{figure}

\begin{deluxetable}{rcrcrcrrr}
\rotate
\tablecolumns{7}
\tablewidth{0pc}
\tablecaption{GRB 091127 Optical Transient Photometry}
\tablehead{

\colhead{Observed} & \colhead{Rest-Frame} & \colhead{}  & \colhead{}  & \colhead{Exp.} & \colhead{}
& \colhead{$m_I$} & \colhead{$m_I$} & \colhead{$M_V$} \\ 

\colhead{Days\tablenotemark{a}} & \colhead{Days\tablenotemark{b}} & \colhead{Magnitude\tablenotemark{c}}  & \colhead{Filter}  & \colhead{Time (s)} & \colhead{Instrument}
& \colhead{OAG+SN\tablenotemark{d}} & \colhead{SN\tablenotemark{e}} & \colhead{SN\tablenotemark{f}}
}
\startdata

0.04263   & 0.02861 & $15.64\pm0.05 $ & UVM2 & 196.6 & \textit{Swift}/UVOT & \nodata & \nodata & \nodata \\
0.04410   & 0.02960 & $ 15.70\pm0.07 $ & UVW1 & 40.9 & \textit{Swift}/UVOT & \nodata & \nodata & \nodata \\
0.06419   & 0.04308 & $16.01\pm0.05 $ & UVW1 & 115.2 & \textit{Swift}/UVOT & \nodata & \nodata & \nodata \\
0.06713   & 0.04505 & $16.26\pm0.13 $ & U & 10.6 & \textit{Swift}/UVOT & \nodata & \nodata & \nodata \\
0.09183 	& 0.06163 & $17.29\pm0.01$ & B & 180 & SMARTS/ANDICAM &  \nodata & \nodata & \nodata \\
0.09183 	& 0.06163 & $17.05\pm0.01$ & V &	120 & SMARTS/ANDICAM &  \nodata & \nodata & \nodata \\
0.09183 	& 0.06163 & $16.76\pm0.01$ & R & 180 & SMARTS/ANDICAM & \nodata& \nodata & \nodata \\
0.09183    & 0.06163 & $16.48\pm0.01$ & I   & 180 & SMARTS/ANDICAM & 16.48 & \nodata & \nodata \\
0.09183 	& 0.06163 & $15.87\pm0.03$ & J  & 180 & SMARTS/ANDICAM & \nodata& \nodata & \nodata \\
0.09183 	& 0.06163 & $15.35\pm0.03$ & H & 120 & SMARTS/ANDICAM & \nodata & \nodata & \nodata \\
0.09183 	& 0.06163 & $14.82\pm0.03$ & K  &180 & SMARTS/ANDICAM &  \nodata & \nodata & \nodata \\
0.10804   & 0.07251 & $16.57\pm0.02  $ & U & 542.9 & \textit{Swift}/UVOT & \nodata & \nodata & \nodata \\
0.13365   & 0.08970 & $ 16.76\pm0.11 $ & U & 25.4 & \textit{Swift}/UVOT & \nodata & \nodata & \nodata \\
0.15803   & 0.10606 & $16.76\pm0.02$ & I & 180 & SMARTS/ANDICAM & 16.77 & \nodata & \nodata \\
0.17632   & 0.11834 & $16.86\pm0.02 $ & U & 819.6 & \textit{Swift}/UVOT & \nodata & \nodata & \nodata \\
0.22483   & 0.15089 & $17.02\pm0.02$ &  I & 180 & SMARTS/ANDICAM & 17.03 & \nodata & \nodata \\
1.15758   & 0.77690 & $19.20\pm0.03$ & I  & 1080 & SMARTS/ANDICAM & 19.25 & \nodata & \nodata \\
1.27663   & 0.85680 & $19.22\pm0.03$ & I  & 1080 & SMARTS/ANDICAM & 19.27 & \nodata & \nodata \\
4.19824   & 2.81761 & $21.24\pm0.06$ & I  & 2160 & SMARTS/ANDICAM & 21.63 & \nodata & \nodata \\
6.15057   & 4.12790 & $21.35\pm0.05$ & I  & 2160 & SMARTS/ANDICAM & 21.79 & 23.96 & $-17.39$ \\
9.17582   & 6.15827 & $21.60\pm0.02$ & I  & 900 & Gemini/GMOS & 22.20 & 23.46 & $-17.89$ \\
9.18961   & 6.16752 & $21.63\pm0.07$ & I  & 2160 & SMARTS/ANDICAM & 22.24 & 23.58 & $-17.78$ \\
11.19510 & 7.51349 & $21.57\pm0.07$ &  I & 2160 & SMARTS/ANDICAM & 22.14 & 22.85 &  $-18.51$  \\
13.16730 & 8.83711 & $21.55\pm0.07$ &  I & 2160 & SMARTS/ANDICAM & 22.11 & 22.60 & $-18.75$ \\
18.05810 & 12.11950 & $21.55\pm0.01$ & I  & 900 & Gemini/GMOS & 22.10 & 22.38 & $-18.97$ \\
20.16250 & 13.53190 & $21.59\pm0.07$ &  I & 2160 & SMARTS/ANDICAM & 22.18 & 22.42 & $-18.93$ \\
22.13420 & 14.85520 & $21.54\pm0.06$ & I  & 2160 & SMARTS/ANDICAM & 22.09 & 22.28 & $-19.07$ \\
24.12250 & 16.18960 & $21.56\pm0.07$ & I  & 2160 & SMARTS/ANDICAM & 22.13 & 22.30 & $-19.05$ \\
47.10720 & 31.61560 & $22.13\pm0.02$ & I  & 900 & Gemini/GMOS & 23.39 & 23.60 & $-17.76$ \\
76.11240 & 51.08210 & $22.31\pm0.01$ & I  & 900 & Gemini/GMOS & 24.14 & 24.33 & $-17.02$ \\
102.03562 & 68.48030 & $22.54\pm0.05$ &  I  & 900 & Gemini/GMOS & \nodata & \nodata & \nodata \\

\enddata

\tablenotetext{a}{Observer frame days after burst trigger at 2009 Nov. 27, 23:25:45.}  
\tablenotetext{b}{Rest frame days after trigger; $t_{rest} = t_{obs} / (1+z)$.}
\tablenotetext{c}{Vega magnitudes corrected for Galactic extinction
  (E$_{B-V} = 0.038$).  In addition to the errors
  quoted, there is a systematic uncertainty of 0.05 mag in the
  SMARTS/Gemini photometric zero-point.}  
\tablenotetext{d}{Observed
  $I$-band magnitude of the OT after subtraction of the
  host galaxy ($I_{\rm host}=22.54$ mag); it is a combination of the
  burst's OAG and SN.}  
\tablenotetext{e}{Observed $I$-band magnitude
  of SN 2009nz.  Both the host galaxy and the OAG have been
  subtracted from the raw observed magnitude of the OT.  The
  brightness of the OAG is modeled by a fit 1--6 days post-burst
  (after host subtraction), with OAG magnitude =$18.944+3.812\times \rm{log}{(t_{\rm obs})}$, where $t_{\rm obs}$ is days
  post-burst in the observer frame (column \#1).}
\tablenotetext{f}{$V$-band absolute magnitude of SN 2009nz, assuming a distance modulus of 42.2 mag and a
  K-correction of $-0.85$ mag.}

\end{deluxetable}

\end{document}